\documentclass[amsmath,amssymb,pre,10pt,nofootinbib]{revtex4-1}
\usepackage{graphicx} 
\usepackage{soul}
\usepackage{upgreek}
\usepackage{hyperref}
\usepackage{xcolor}
\usepackage{orcidlink}
\usepackage{perpage}
\usepackage{amsmath}
\usepackage{hyperref}
\usepackage{amsfonts}
\usepackage{amssymb}
\usepackage{amsbsy}
\usepackage{mathrsfs}
\usepackage{comment}
\usepackage{amsthm}
\usepackage{graphicx}    
\usepackage{rotating}    
\usepackage{epsfig}
\usepackage{color}       
\usepackage{xcolor}      

\definecolor{gruen}{rgb}{0,0.625,0}     
\definecolor{rot}{rgb}{0.75,0,0}        
\definecolor{blau}{rgb}{0,0,0.75}       

\newcommand{\BEQ}{\begin{equation}}     
\newcommand{\BEA}{\begin{eqnarray}}
\newcommand{\BD}{\begin{displaymath}}
\newcommand{\EEQ}{\end{equation}}       
\newcommand{\EEA}{\end{eqnarray}}
\newcommand{\ED}{\end{displaymath}}
\newcommand{\vep}{\varepsilon}          
\newcommand{\D}{{\rm d}}                
\newcommand{\demi}{\frac{1}{2}}         
\newcommand{\lap}[1]{\overline{#1}}     

\renewcommand{\vec}[1]{\boldsymbol{#1}} 

\newcommand{\appsection}[2]{\setcounter{equation}{0}\setcounter{subsection}{0}\setcounter{table}{0}\setcounter{figure}{0}
\section*{Appendix #1. #2}
\renewcommand{\theequation}{#1.\arabic{equation}}
              \renewcommand{\thesection}{#1}\renewcommand{\thetable}{#1\arabic{table}}\renewcommand{\thefigure}{#1.\arabic{figure}} }

\begin{document}
\title{Bio-heat regimes in fractal-based models of tumors}
\author{S\'ebastien Fumeron}
\affiliation{Laboratoire de Physique et Chimie Th\'eoriques (CNRS UMR 7019),\\  Universit\'e de Lorraine Nancy,
B.P. 70239, F -- 54506 Vand{\oe}uvre l\`es Nancy Cedex, France}
\author{Malte Henkel}
\affiliation{Laboratoire de Physique et Chimie Th\'eoriques (CNRS UMR 7019),\\  Universit\'e de Lorraine Nancy,
B.P. 70239, F -- 54506 Vand{\oe}uvre l\`es Nancy Cedex, France}
\affiliation{Centro de F\'{i}sica Te\'{o}rica e Computacional, Universidade de Lisboa, \\Campo Grande, P--1749-016 Lisboa, Portugal}
\author{Alexander L\'opez}
\affiliation{Escuela Superior Polit\'ecnica del Litoral, ESPOL,\\ 
Departamento de F\'isica, Facultad de Ciencias Naturales y Matem\'aticas,\\ Campus Gustavo Galindo
 Km. 30.5 V\'ia Perimetral, P. O. Box 09-01-5863, Guayaquil, Ecuador}
 \affiliation{GISC, Departamento de F\'{\i}sica de Materiales, Universidad Complutense, E-28040 Madrid, Spain}
\begin{abstract}
Anomalous heat diffusion is investigated for biological tissues displaying a fractal structure and long-term thermal memory, 
which is modelled via a fractional derivative. For increasing values of the fractional derivation order, the tissue temperature displays three kinds of bio-heat regimes: 
damped (or sub-diffusive), critical damping and under-damped oscillations. 
The temperature profiles depend on the fractal dimension of the tissue but notably also on a parameter related to its topology: 
the spectral dimension. The parametric analysis reveals that these two parameters have antagonistic effects on the pseudo period of the temperature oscillations and their
amplitudes. We discuss how our results might impact some treatment protocols.
~\\[3.0cm]
\end{abstract}

\maketitle


\setcounter{footnote}{0}

\section{Introduction} \label{sec1}
 
Cancer has now become one of the major causes of mortality worldwide and hence, 
the biophysics of cancer is currently raising lots of attention as it offers novel insights into developing new models of tumor growth, 
better diagnostic tools and more efficient therapies. In particular, 
cell nuclei are rich in proteins involved in cellular homeostasis and therefore, targeted overheating has been proven 
to be a potent but challenging tool to treat cancer. In the case of mild hyperthermia, malignant cells are exposed to supra-physiological temperatures (39-43$^o$C) 
to induce the direct apoptosis of tumors and sensitise 
them to radiation and anticancer drugs. Yet at the same time, incomplete hyperthermia may trigger the upregulation in expression of heat-shock proteins, 
which are anti-apoptotic and result in thermo-tolerance \cite{tomasovic1983heat,fortin2000overexpression,calderwood2002targeting,kassis2021heat}. 
This latter point strongly limits the frequency of thermotherapy sessions which must be spaced several days apart.

Hence, the main scientific challenge with thermotherapy is to perfectly control temperature gradients not only at the edges of the neoplastic tissues 
(the thermal dose must be delivered to the tumor only and spare surrounding healthy cells), 
but also within them: indeed the existence of highly vascularised regions within a tumor (angiogenesis) 
can result in local "cool spots" which may jeopardize the efficiency of the treatment. 

Models investigating heat transfer in biological matter started with the pioneering works of C.K. Pennes in 1948 \cite{pennes_analysis_1948}. 
The so-called Pennes bio-heat equation is based on a local enthalpy balance in the presence of heat conduction (Fourier's law) 
and heat exchange between the tissue and the bloodstream:
\begin{equation} 
\rho_t c_t \frac{\partial T}{\partial t} = \lambda \nabla^2 T- \rho_b \omega_b c_b (T - T_{\text{b}})+ Q_{\text{meta}} \label{pennes}
\end{equation}
where $T=T(t,r)$ is the local temperature in the tissue, $t$ is time, $r=|\vec{r}|$ is space with rotation-invariance implicitly admitted, 
$\nabla^2$ is the Laplace operator, $\rho_t$ is the tissue density, $c_t$ is the tissue specific heat, 
$\lambda$ is the mean thermal conductivity, 
$Q_{\text{meta}}$ the metabolic heat generation, $\rho_b$ the blood density, $c_b$ the specific heat of the blood, $\omega_b$ is the blood perfusion rate and finally 
$T_{\text{b}}$ the arterial blood temperature. Pennes' original contribution lies in the phenomenological perfusion term which assumes capillary-bed equilibration: 
the venous blood and the tissue are at thermal equilibrium \cite{hristov2019bio}. 
Despite early successes in therapeutic hyperthermia, tissue perfusion by heat clearance methods, 
and whole human body thermal models 
under conditions of environmental stress (see \cite{charny1992mathematical} and references therein), 
the ordinary Pennes equation suffers from severe limitations when it comes to tumors: 
in particular neither the thermo-tolerance nor the morphological characteristics of malignant tissues enter in (\ref{pennes}). 

As a matter of fact, carcinogenesis entails architectural changes in the tissue structures, 
as a result of a delicate interplay between mechanical stresses and biochemical pathways. 
For instance, in the case of human breast carcinoma, 
tumor cells realign bundles of collagen perpendicular to the tumor boundary to facilitate local invasion: 
this is known as tumor-associated collagen signature \cite{provenzano2006collagen, conklin2011aligned}. 
Another morphological biomarker of cancer is the fractality of tissues. Since the pioneering works of Mandelbrot \cite{mandelbrot1983fractal}, 
fractal geometry turned out to be an unescapable tool in physics, 
allowing to unify random walks \cite{klafter1996beyond}, critical phenomena \cite{saberi2015recent,Henkel08,BoettcherHerrmann2021}, 
geology \cite{turcotte1989fractals,Sornette06}, quantum mechanics \cite{kroger2000fractal} or turbulence \cite{sreenivasan1991fractals}, 
within a common framework. In biophysics, fractal analysis has proven to be a potent tool for assessing tumor aggressiveness from histological data, 
in the case of breast cancer 
\cite{tambasco2008relationship,hermann2015fractal,chan2016automatic,elkington2022fractal}, colorectal cancer 
\cite{esgiar2002fractal,cusumano2018fractal,elkington2022fractal}, lung cancer \cite{lennon2015lung}, 
pancreatic cancer \cite{scampicchio2016assessment,elkington2022fractal} and 
prostate carcinoma \cite{de2013quantification,waliszewski2016quantitative,elkington2022fractal}. 

In this article we propose a new bio-heat model which takes into account the thermo-tolerance as a memory effect via temporal fractional derivatives. 
For this purpose, we use the Caputo fractional derivative which encompasses a convolution of the time derivative with a power law kernel that provides the required non-locality in time. 

This work is organized as follows: in section~\ref{sec2}, we introduce our new bio-heat equation that encompasses thermo-tolerance effects 
as well as the fractal structure of neoplastic tissues. Its analytic solution via a separation of variables is outlined. 
In section~\ref{sec:3}, the explicit solution is given and the thermal answer of biological tissues and their high sensitivity to both 
their fractal and spectral  dimensions is analysed. Finally, a parametric study is performed which suggests pathways to optimize mild hyperthermia protocols.  
We summarise our results in section~\ref{sec4}. 
Three appendices contain details of calculation, background on the (Caputo) fractional derivative and further numerical data. 

\section{Fractal Pennes' Model with memory effect} \label{sec2}

A description of many biological tissues via fractal media is often accurate and has become common. 
A possible explanation or motivation may lie in the avascular tissue growth mechanisms \cite{ribeiro2017fractal}. Within a population of competing cells, each cell must 
move to occupy space in order to optimise its access to nutrients: 
the fractal geometry then emerges spontaneously as the mean to minimize the competitive pressure between cells when their number becomes large. 
Another possible explanation was provided by tissue formation mechanisms \cite{giorgio2015differential,leggett2019motility}: 
multicellular collectives first random-walk and then assemble into fractal-like clusters driven by diffusion-limited aggregation. 
Fractality may also arise from the homeostatic functioning of a tissue, which results from multi-scale cross-interacting processes 
(upward microscopic causation due to DNA, proteins \ldots and downward causation due to epigenetic mechanisms, 
signalling pathways \ldots) \cite{waliszewski2001relationship}. 
Fractality may finally originate in specific biological functions: 
for instance, to maximize the exchange surface within the volume of the torso, 
organs such as intestine or lungs display a space-filling multi-scale structure 
({\it alveoli}, {\it villi} \ldots) that identifies with a fractal.

We begin with some background information on fractals: it is known that
the description of self-similar structures requires at least the definition of three dimensionless numbers \cite{rammal1983random}: 
the embedding euclidean space dimension $d$ (an integer), the fractal dimension $D_f$ and the spectral dimension $d_s$. These in general  
satisfy the double inequality $d_s\leq D_f\leq d$ \cite{rammal1983random} 
(recently, it was suggested that six independent numbers were needed to encompass the topological features of fractals \cite{patino2024morphological}). 
The {\em fractal dimension} $D_f$ \cite{Strogatz94,Sornette06,BoettcherHerrmann2021} 
can be defined from the scaling law that relates the resolution (length scale $\varepsilon$) to the number of boxes 
$N(\varepsilon)\sim \vep^{-D_f}$ required to cover the fractal (Minkowski-Bouligand or `box-counting' dimension): 
\begin{equation}
 D_f=-\lim_{\varepsilon\rightarrow0}\frac{\ln N(\varepsilon)}{\ln \varepsilon}.   
\end{equation}
It is generally a non-integer quantity which measures the degree of filling of the embedding space: 
for instance, the celebrate {\em Peano curve} is a full space-filling object of fractal dimension $D_f=2=d$. 
A simple example of a non space-filling fractal curve is the {\em von Koch curve} with a non-integer 
$D_f=\frac{\ln 4}{\ln 3}\approx 1.26$ \cite{Strogatz94,Sornette06,Henkel08,BoettcherHerrmann2021}. 
The fractal dimension $D_f$ has been identified as a morphometric biomarker in oncopathology. 
Histological cuts from breast, cervix, colon, lung, pancreas and prostate revealed two important features of tumors compared to healthy tissues. 
\begin{enumerate} 
\item the fractal dimension $D_f$  of a tumoral tissue is higher than that of the corresponding normal tissue 
\cite{wu2008medical,jayalalitha2011fractal,elkington2022fractal}. 
\item the fractal dimension $D_f$ increases with the cancer grade \cite{tambasco2008relationship,elkington2022fractal}. 
\end{enumerate}
These two properties are correlated to the fact that malignant tumors fill space to invade the surrounding tissues 
(but note that the inverse correlation was found in \cite{chan2016automatic}, as a result of different stains used to prepare histology specimens).

It should not come as a surprise that $D_f$ is not enough to fully characterise a fractal: the Peano curve and the Mandelbrot fractal have the same fractal dimension $D_f=2$, 
but do not share the same topology. The role of topology is encompassed within the 
{\em spectral dimension} $d_s$, which depends on topological properties of the
fractal (such as connectivity) \cite{rammal1983random}. Physically, $d_s$ is a crucial parameter in random walk processes \cite{alexander1982density,o1985diffusion}: 
if $S_N$ is the average number of distinct sites visited on a fractal lattice during an \textit{N}-step random walk, 
then the spectral dimension is the scaling exponent such that \cite{rammal1983random}
\begin{eqnarray}
    &&S_N\sim N^{d_s/2}\;\;\;\text{if}\;\;d_s<2 
\end{eqnarray}
Let us give some typical examples: for a $2D$ percolation cluster or a corrosion front, 
the Alexander-Orbach conjecture states that $d_s=4/3$ independently of the Euclidean dimensionality, for $d\ge 2$ \cite{alexander1982density}). 
For a $2D$ Sierpinski gasket, it was found that $d_s=2\frac{\ln 3}{\ln 5}\simeq 1.365$ \cite{alexander1982density}, 
whereas $D_f=\frac{\ln 3}{\ln 2}\simeq 1.585$. Experimentally, Brillouin scattering measurements provided spectral dimension 
$d_s= 1.252\pm 0.061$ in silica aerogels \cite{courtens1987brillouin}. 

Therefore, modelling the diffusion of heat on a fractal domain in an embedding space of dimension $d$ (later we shall take $d=2$, the dimension of an histological slice) 
requires to replace the Laplace operator term appearing in (\ref{pennes}) by \cite{o1985analytical,o1985diffusion} 
\begin{equation} 
    \lambda \nabla^2 T \;\;\longrightarrow\;\; 
    \frac{\lambda}{r^{d-1}}\frac{\partial}{\partial r}\left( r^{d-1}\frac{\partial T}{\partial r}\right)\;\;\longrightarrow\;\;
    \frac{1}{r^{D_f-1}} \frac{\partial}{\partial r}\left(\lambda r^{D_f-1-\theta}\:\frac{\partial T}{\partial r}\right) 
    \label{transfo-fractal}
\end{equation}
where we go over, from a space-filling geometry of dimension $d$, to a fractal tissue with fractal dimension $D_f$. In our context, we must have $1<D_f\leq d=2$. In addition, again following \cite{o1985analytical,o1985diffusion}, a further modification of the Laplace operator is required, which uses the exponent of anomalous diffusion 
\begin{equation} \label{gl:4}
    \theta=\frac{2D_f}{d_s}-2>0
\end{equation} 
The parameter $\theta$ is related to the spectral dimension $d_s$, such that the lower bound $\theta=0$ is only reached in the Euclidean case 
(note that the dimensionality of the thermal conductivity appearing in (\ref{transfo-fractal}) is now 
$W.m^{\theta-1}.K^{-1}$).\footnote{It can be shown that the form (\ref{transfo-fractal}) of the laplacian is required in $d$-dimensional diffusion in order to reproduce the 
sub-diffusive behaviour of an averaged spatial distance $\langle r^2(t)\rangle\sim t^{2/(2+\theta)}$ \cite{o1985diffusion}.}  
As far as we know, current histological data do not provide the spectral dimension and hence the forthcoming results will be parametrised by the couple 
$\{D_f,d_s\}$.

Another major problem when considering hyperthermia lies in thermoresistance effects. 
When a tumor undergoes a heat stress, the signalling pathways involved in heat shock proteins (HSPs) 
production are triggered in order to prevent protein unfolding or aggregation, 
and to help repair misfolded or denatured proteins \cite{subjeck1982heat}. 
This protective cellular response is responsible for a transient enhanced tolerance of the tumor to subsequent thermal stresses. 
Thus, to account for this memory effect, expressed as a non-locality in time, 
one replaces the ordinary first-order time derivative by a Caputo fractional derivative of non integer order $\alpha$ ($n-1<\alpha<n$, $n\in\mathbb{N}$)
\begin{equation}
\frac{\partial T}{\partial t}(t)\;\;\;\;\longrightarrow\;\;\;\;{^c}D_t^\alpha T(t)=\frac{1}{\Gamma(n-\alpha)}
\int_{t_0}^t\!\D u\: \frac{1}{(t-u)^{\alpha+1-n}} \frac{\D^n T(u)}{\D u^n}, \label{switch-caputo}
 \end{equation}
which involves the ordinary $n^{\text{th}}$-order derivative $T^{(n)}(t)=\frac{\D^n T(t)}{\D t^n}$. This corresponds to a power law singular kernel at the end point, 
i.e. short-time events have the largest impact on the system dynamics. 
The parameter $t_0$ is the characteristic remembrance time, and it will be chosen as $0$ in the remainder of this work. 
The $\alpha$ of the fractional derivative represents the strength of the memory: 
the larger $\alpha$ is, the slower the system forgets \cite{du2013measuring}. The Caputo derivative is chosen among the many existing fractional derivatives 
as it is particularly well-suited to deal with Cauchy initial-value problems \cite{kilbas2004cauchy,Diethelm2010}. 
We shall consider separately the two cases $n=1$ and $n=2$ to explore the memory effects in the diffusive and super diffusive regimes. 
Appendix~B collects properties of $^cD^{\alpha}_t$ needed in this work. 

Hence, the equation governing bio-heat transfer inside a fractal tissue with a thermal memory effect becomes
\begin{equation} 
\rho_t c_t \tau^{\alpha-1} {^c}D_t^\alpha T 
= \frac{1}{r^{D_f-1}}\frac{\partial}{\partial r}\left(\lambda r^{D_f-1-\theta}\:\frac{\partial T}{\partial r}\right)
- \rho_b \omega_b c_b (T - T_{\text{b}})+ Q_{\text{meta}} 
\label{fractal-pennes}
\end{equation}
and will be taken as our starting point for the present study. At thermal equilibrium, the homeostatic body-temperature $T_0$ is given by
\BEQ
T_0=T_b+ \frac{Q_{\text{meta}}}{\rho_b \omega_b c_b}
\EEQ
and the evolution described by (\ref{fractal-pennes}) will relax to this value. 
The parameter $\tau$ in eq.~(\ref{fractal-pennes}) guarantees the dimensional consistency. 
To achieve a more compact form of (\ref{fractal-pennes}), we define the following parameters as short-hands:
\begin{equation} \label{gl:racc}
\mathcal{D} = \frac{\lambda}{\rho_t c_t \tau^{\alpha-1}} \;\; , \;\; 
\mathcal{D} \gamma = \frac{\rho_b c_b \omega_b}{\rho_t c_t \tau^{\alpha-1}}. \;\; \end{equation}
Using the transformation
$T \mapsto \tilde{T}=T -T_0$ to recover a homogeneous equation, we are finally led to 
\begin{equation} 
{^c}D_t^\alpha \tilde{T} = 
\frac{\mathcal{D}}{r^{D_f-1}}\frac{\partial}{\partial r}\left(r^{D_f-1-\theta}\:\frac{\partial \tilde{T}}{\partial r}\right)- \mathcal{D}\gamma \tilde{T} 
\label{homog-pennes}
\end{equation}
We note that this equation introduces a natural velocity which scales with the relaxation time as 
\begin{equation}
    V=\mathcal{D}^{\frac{1}{\theta+2}}\tau^{\frac{\alpha}{\theta+2}-1}\sim \tau^{\frac{1}{\theta+2}-1}
\end{equation}
This means that higher values of $\theta$ are associated with slower bio-heat phenomena. 

\begin{figure}[tb]
\begin{center}
\includegraphics[width=.46\hsize]{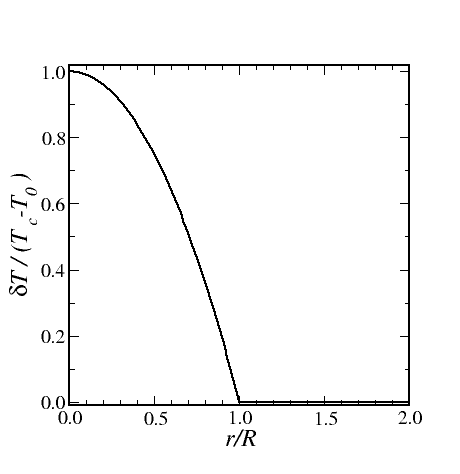} 
\end{center}
\caption[fig1]{\small Form of the initial profile $\delta T = T(0,r)-T_0$. 
\label{fig1} 
}
\end{figure}

The fractal Pennes equation (\ref{homog-pennes}) can now be solved by means of the separation ansatz
$\tilde{T}(t,r) = f(t) g(r)$, generalising the standard treatment of the fractional diffusion equation \cite{manapany2024}. 
This leads to the following set of independent equations 
\begin{subequations}
\begin{align}
{^c}D_t^\alpha f(t)&= -\omega^\alpha f(t) \label{sep-time} \\
\frac{1}{r^{D_f-1}}\frac{\partial}{\partial r}\left(r^{D_f-1-\theta}\:\frac{\partial g(r)}{\partial r}\right)&= - \left(\frac{\omega^\alpha}{\mathcal{D}}-\gamma\right) g(r) 
\label{sep-space}
\end{align}
\end{subequations} 
The eigensolutions of the first equation (\ref{sep-time}) is obtained from the Mittag-Leffler function $\text{E}_{\alpha}$, see \cite{Diethelm2010,Gorenflo} and appendix~B, 
according to\footnote{As it stands, this merely holds for $0<\alpha\leq 1$, but we anticipate the further initial condition (\ref{init2-bis}) below 
which extends the range of validity to $0<\alpha\leq 2$.} 
\begin{equation}
f(t) = f_{\omega^{\alpha}}(t)=f_0\; \text{E}_{\alpha}\left(-\omega^\alpha t^{\alpha}\right) \label{eigen-time} 
\end{equation}
which also depends on the separation constant $\omega^{\alpha}$. Here $f_0$ is a constant that would be determined through  initial conditions that are defined below. 
The second equation (\ref{sep-space}) is solved by changing the dependent variable to $g(r)=r^{\beta}z(r)$ with 
\BEQ \label{gl:beta}
\beta=\theta+3-D_f
\EEQ 
Then, after some algebraic manipulations (see appendix~A), 
we find that  the eigensolution to (\ref{sep-space}) becomes 
\begin{equation} \label{eq13}
    g(r)= r^\frac{\beta-1}{2}\left[D_1  J_{(\beta-1)/(\theta+2)}\left(\frac{2\sqrt{\gamma_{\alpha}}}{(\theta+2)} r^{(\theta+2)/2}\right) 
+  D_2  J_{-(\beta-1)/(\theta+2)}\left(\frac{2\sqrt{\gamma_{\alpha}}}{(\theta+2)} r^{(\theta+2)/2}\right) \right]
\end{equation}
with $\gamma_{\alpha}=\omega^\alpha/\mathcal{D}-\gamma$ and undetermined constants $D_{1,2}$. 
Combining this with (\ref{eigen-time}) and summing over all possible solutions finally gives the temperature field
\begin{eqnarray}
    \lefteqn{T(t,r) = T_0+\int\!\D \omega\: c_{\omega}\: r^{{\beta-1}/{2}}\, \text{E}_{\alpha}\left(-\omega^\alpha t^{\alpha}\right)\times} \nonumber\\
    && \times\left[D_1  J_{(\beta-1)/(\theta+2)}\left(\frac{2\sqrt{\gamma_{\alpha}}}{(\theta+2)} r^{(\theta+2)/2}\right) 
+  D_2  J_{-(\beta-1)/(\theta+2)}\left(\frac{2\sqrt{\gamma_{\alpha}}}{(\theta+2)} r^{(\theta+2)/2}\right) \right] \label{general-sol}
\end{eqnarray}
where the integral/sum runs over all admissible values of $\omega$ which will be found, along with the constants $c_{\omega}$, from the initial condition.

As initial and/or boundary conditions, we assume the temperature-profile
\begin{subequations}  \label{init2-total}
\begin{equation} \label{init2}
   T(t,R)=T_0 \;\; , \;\; 
   T(0,r)=  (T_c-T_0) \left(1 - \frac{r^2}{R^2}\right)\Theta(R-r) +T_0 \;\; , \;\; 
   \left\{ \begin{array}{c} \left(\frac{\partial T}{\partial r}\right)(t,0)=0 \\[0.1cm]
                           \mbox{\rm\small $T(t,r)$ has a local maximum at $r=0$} 
          \end{array} \right. 
\end{equation}
with $R$ the attenuation length and $T_c$ being the temperature at the centre of the tissue ($\Theta$ denotes the Heaviside function). 
This is illustrated in figure~\ref{fig1}. 
The first condition (\ref{init2}) is introduced to recover the body temperature far enough from the heated region. 
The second condition (\ref{init2}) corresponds to the initial temperature profile resulting from the heating procedure 
(which we admit since it will lead to an explicit analytic solution in section~\ref{sec:3}). 
The last constraint guarantees that the temperature (\ref{general-sol}) and the heat flux $-\partial_r T(t,r)$ are continuous at the centre of the domain. 

The initial and boundary conditions (\ref{init2}) completely determine the solution of (\ref{homog-pennes}) for the fractional index $0<\alpha\leq 1$. 
But if $1<\alpha\leq 2$, a further initial condition is needed, which we shall take as
\BEQ \label{init2-bis}
\frac{\partial}{\partial t} T(0,r) = 0, 
\EEQ
\end{subequations}
which means that the initial configuration specified in (\ref{init2}), see again figure~\ref{fig1}, is assumed to be initially at rest. 
This would be consistent with the consideration of initial homeostasis. The combined initial conditions 
(\ref{init2},\ref{init2-bis}) then will fix the unique solution of (\ref{homog-pennes}) for all $0<\alpha\leq 2$ \cite{manapany2024,Diethelm2010,Gorenflo}. 


\section{Temperature-profiles in heated tissues} \label{sec:3}

\begin{figure}
    \centering
    \includegraphics[width=0.5\linewidth]{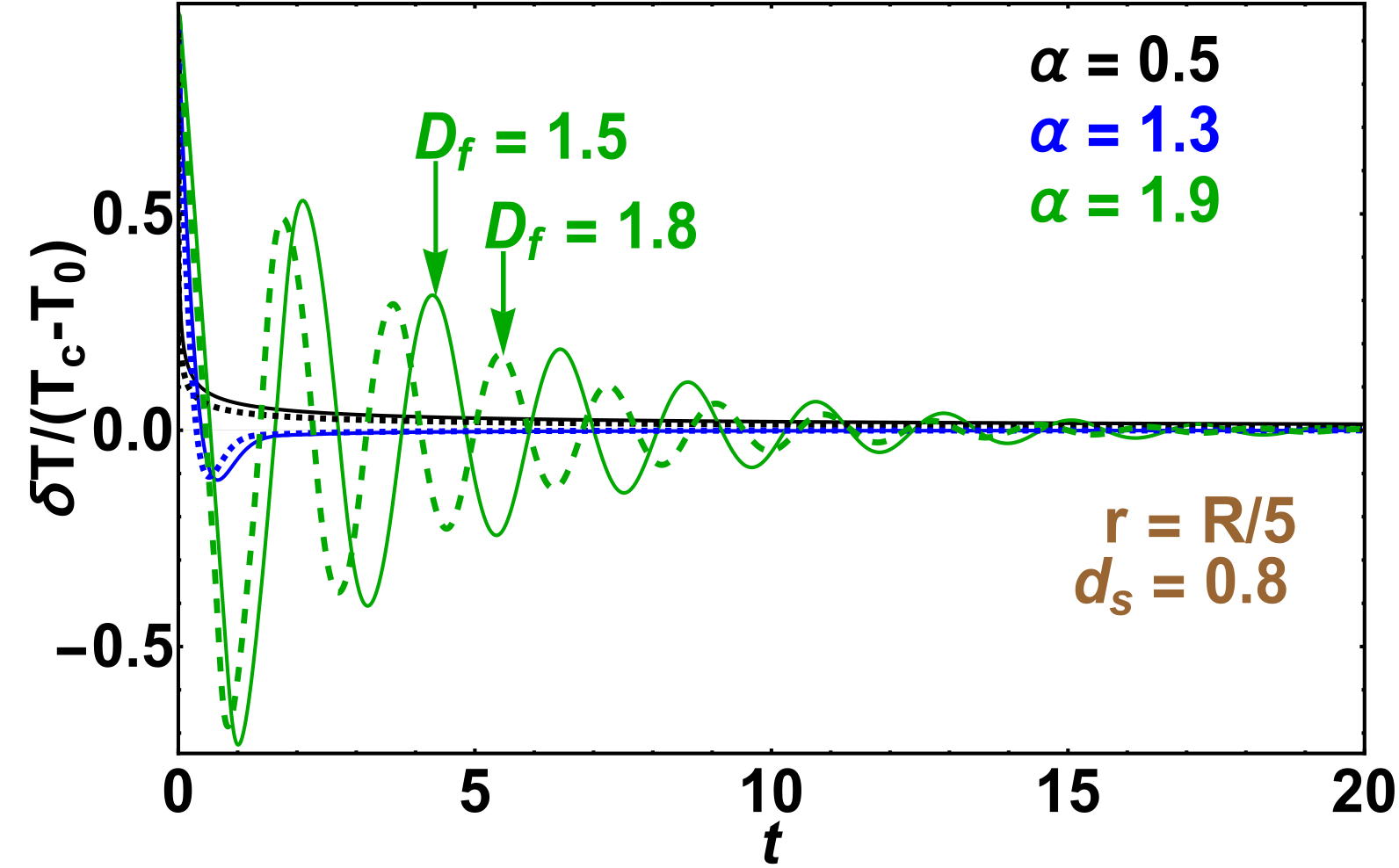}~\includegraphics[width=.5\linewidth]{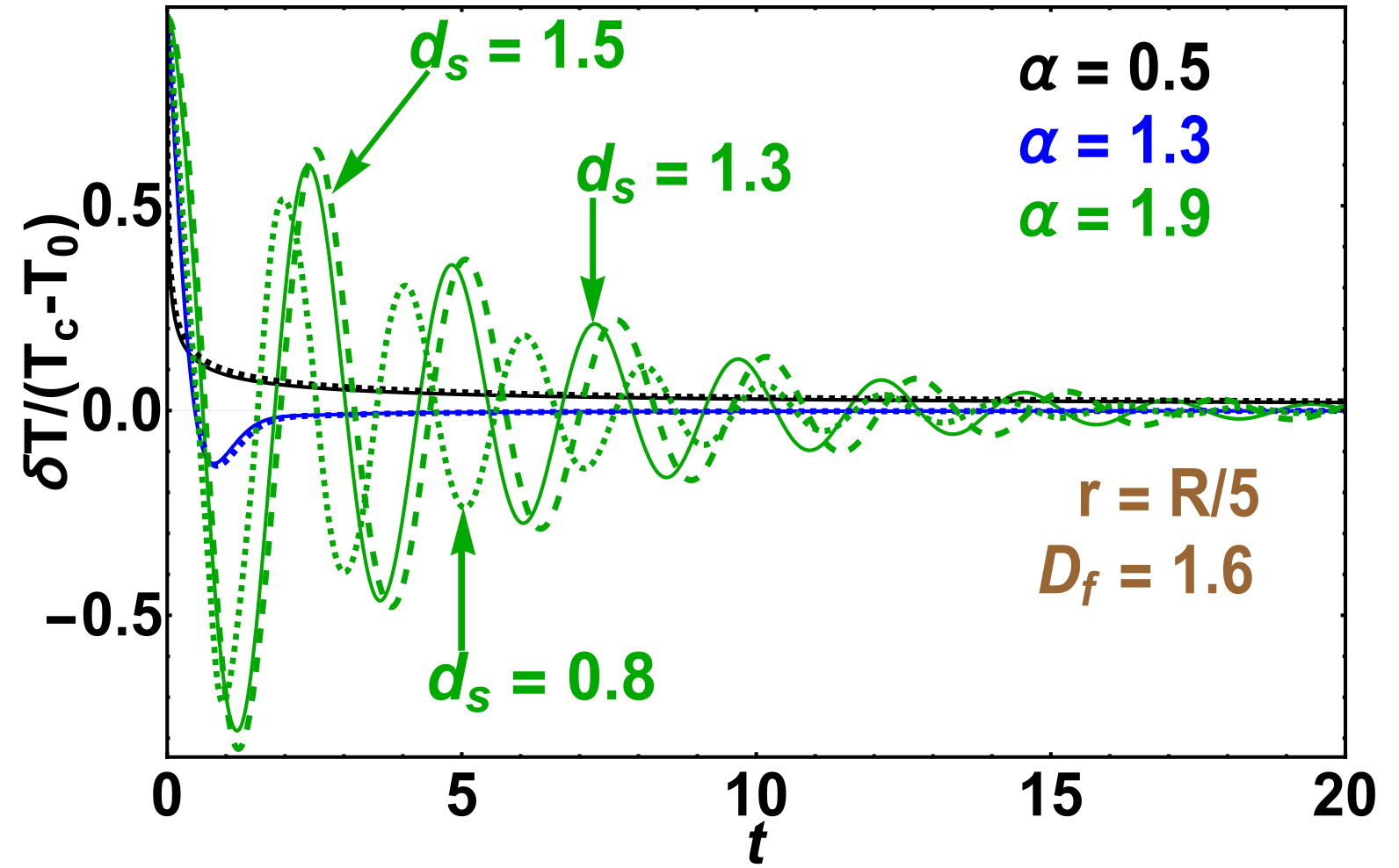}
    \caption{Time evolution of the normalised differential temperature at $r=0.2R$, at three selected values for $\alpha=0.9, 1.3$ and $1.9$. 
    The left panel corresponds to fixed spectral dimension and the thin (thick) lines is parametrised by the fractal dimension  $D_f=1.5$ or $D_f=1.8$. 
    In the right panel, the fractal dimension is set to $D_f=1.6$ and the three different lines correspond to spectral dimensions $d_s=0.8$, $d_s=1.3$ and $d_s=1.5$ as depicted by the arrows.}
    \label{fig:6}
\end{figure}

The formal solution (\ref{general-sol}) discussed above must satisfy some consistency conditions. 
For $0<\alpha\leq 2$, only $\beta>2$ is physically acceptable, since the other possibility $\beta=1$ would imply the fractal dimension satisfies $D_f>2$, 
which is non-physical as we are considering two-dimensional histological models. 
The solutions for $1<\beta<2$ are in turn discarded because of the boundary conditions of relevance for actual implementations of the model. 
This leads to an additional inequality relating $D_f$ and $d_s$
\begin{equation}
    \frac{d_s}{2-d_s}<D_f \label{ineq}
\end{equation}
Moreover, for $\beta>2$ the boundary condition ${T}(t,R)=T_0$ allows us to determined the frequency spectrum, which is explicitly given by 
\begin{equation}
\omega^\alpha_n=\mathcal{D}\gamma+\mathcal{D}\left(\frac{\kappa_{-1+d_s/2,n}\, D_f/d_s}{R^{2D_f/d_s}}\right)^2
\label{sign-arg}
\end{equation}
%
where $\kappa_{-\nu,n}$ is the $n^{\rm th}$ zero of the Bessel function $J_{-\nu}(x)$ such that $J_{-\nu}(\kappa_{-\nu,n})=0$ \cite{Abra65}. 
Therefore, since $\nu=(\beta-1)/(\theta+2)=1-D_f/(\theta+2)=1-d_s/2$, the temperature profile, satisfying all initial and boundary conditions (\ref{init2-total}), 
reads for $0<\alpha\leq 2$
\begin{eqnarray} \label{glTV:32}
    T(t,r) = T_0+\left(\frac{r}{R}\right)^{D_f/d_s-D_f/2}\sum_{n=1}^\infty \: \, d_{1-d_s/2,n}\,\text{E}_\alpha(-\omega_n^\alpha t^\alpha) 
     J_{-1+d_s/2}\left(\kappa_{-1+d_s/2,n}\left(\frac{r}{R}\right)^{D_f/d_s}\right) \label{solutionT}
\end{eqnarray}
where the expansion coefficients read as
\begin{equation} \label{glT:31}
d_{\nu,n}=\frac{2(T_c-T_0)}{J^2_{1-\nu}(\kappa_{-\nu,n})}\left[\frac{J_{1-\nu }(\kappa_{-\nu,n})}{\kappa_{-\nu,n}}
-\frac{2^{\nu } \kappa_{-\nu,n}^{-\nu } \, 
_1F_2\left(\frac{s}{2}-\nu +1;1-\nu ,\frac{s}{2}-\nu +2;-\frac{\kappa_{-\nu,n}^2}{4}\right)}{(-2 \nu +s+2)\,\Gamma (1-\nu ) }\right]  
\end{equation}
with $_pF_q(a;b;z)$ being the hypergeometric function and $s=2d_s/D_f$. The most remarkable feature about (\ref{solutionT}) 
is that the order of the Bessel functions, their zeros and the expansion coefficients only depend on the spectral dimension but not on $D_f$. 
This latter only appears as the power law for the reduced radius and in the expression of the pulsation (\ref{sign-arg}). 
The exponent of anomalous diffusion that naturally arises in the Laplace operator is not a relevant parameter in our model. 
The detailed calculations proving (\ref{solutionT}) and (\ref{glT:31}) are provided in appendix~A. 

\begin{figure}
    \centering
    \includegraphics[width=1\linewidth]{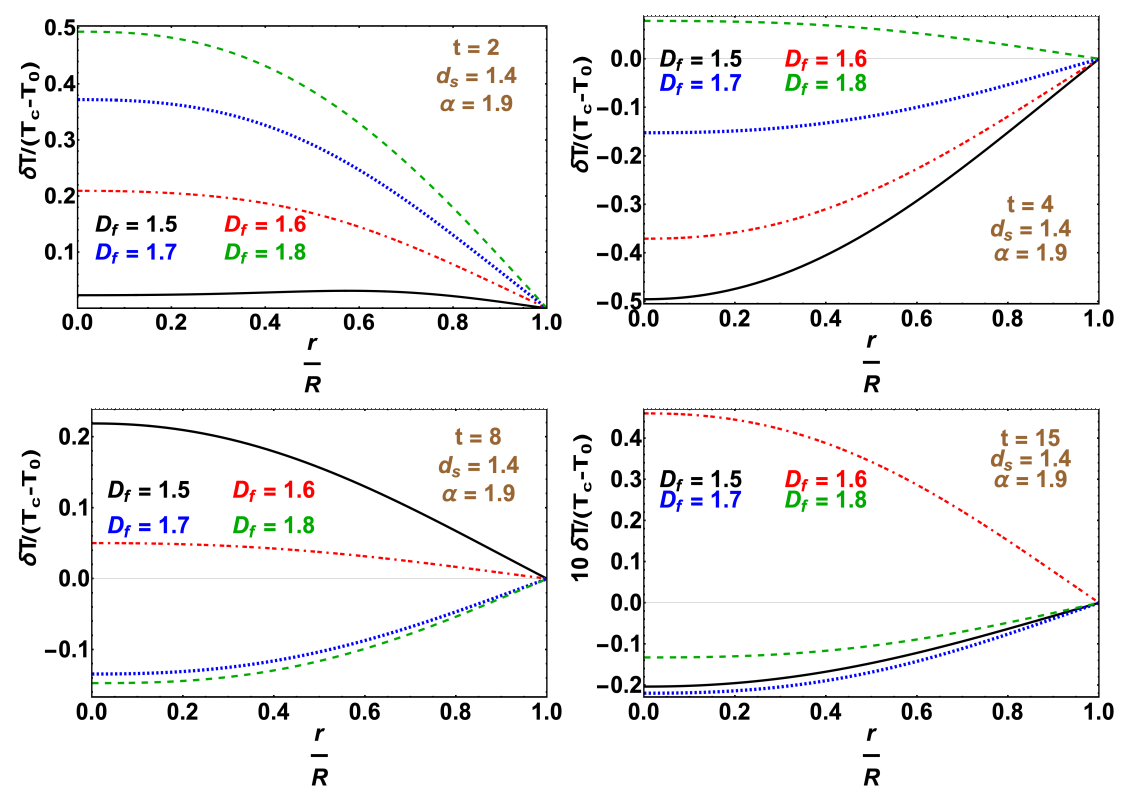}
    \caption{Snapshots of the normalised differential temperature profile versus normalised distance $r/R$, at four selected times for $\alpha=1.9$ and $d_s=1.4$. 
    The curves are parametrised by the fractal dimension $D_f$.}
    \label{fig:4}
\end{figure}

\begin{figure}
    \centering
    \includegraphics[width=1\linewidth]{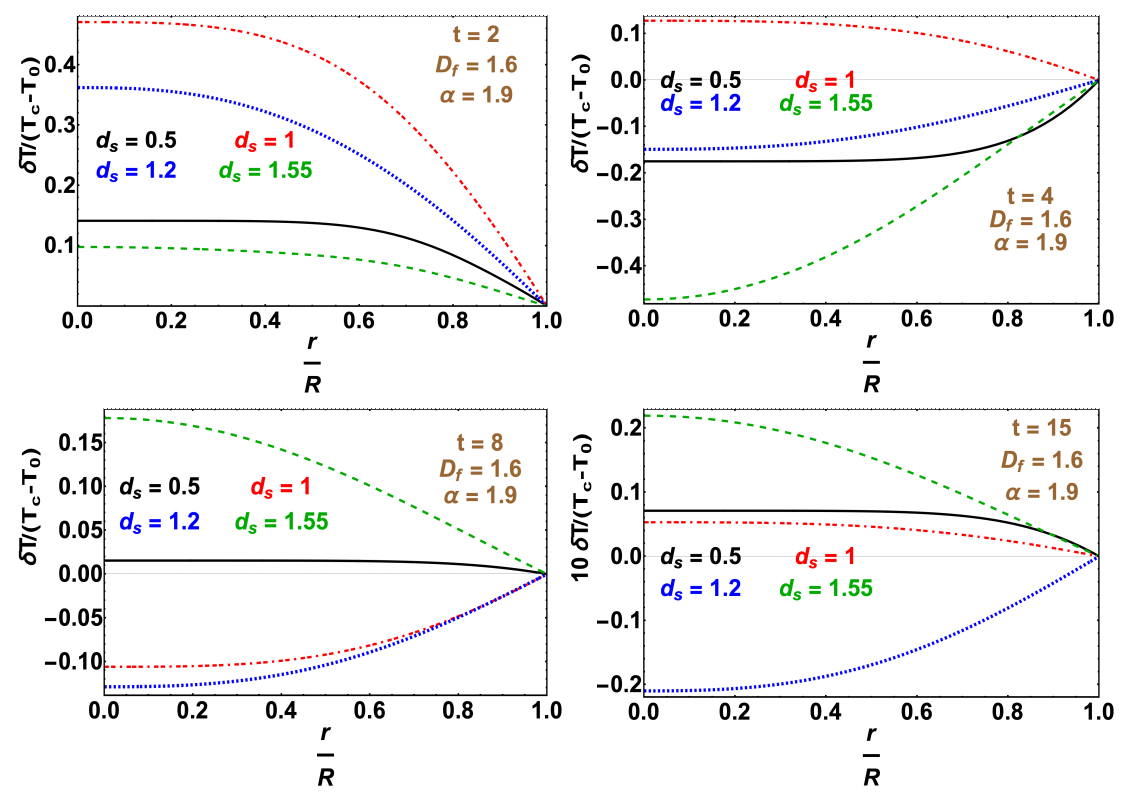}
    \caption{Snapshots of the normalised differential temperature profile versus normalised distance $r/R$, at four selected times for $\alpha=1.9$ and $D_f=1.6$. 
    The curves are parametrised by the spectral dimension $d_s$.}
    \label{fig:3}
\end{figure}

In what follows, we study the resulting thermal profiles. 
First, we consider the time-dependence of the enveloppe, at a particular point $r=0.2 R$. 
In particular, we are interested in the consequences of changing the order $\alpha$ of the time derivative. 
In the left panel of figure~\ref{fig:6}, we depict the temperature-field in the tissue for different values of time derivative order $\alpha$ 
and of the fractal dimension $D_f$ (the spectral dimension is fixed to $d_s=0.8$). 
First, we notice the existence of three different regimes for increasing values of $\alpha$. 
For $\alpha<1$, the bio-heat response follows a sub-diffusive regime: the tissue temperature decays monotonously and it stays above the homeostatic temperature. 
The damping is all the more swift that the fractal dimension is high. For $\alpha=1.3$ (only one zero for the Mittag-Leffler function, see table 7.4 in \cite{Diethelm2010}), 
we observes a first kind of super-diffusive regimes consisting in critically damped oscillations: as before, they occur earlier for higher $D_f$. For high values of $\alpha$, 
there is a second super-diffusive regime consisting in under-damped oscillations. Two trends can be extracted: 
(1) the apparent period of these oscillations increases with time, and (2) the lower $D_f$, the larger the apparent period and the amplitude of the oscillations. 
These observations are confirmed for values of $D_f$ close to the integer values 1 and 2, and they are in qualitative agreements with the trends analysed in (\cite{manapany2024}). 
To connect these results with hyperthermia, it means that in this regime, the tissue will undergo cycles of hot-cold thermal stresses.

The right panel of figure~\ref{fig:6}  depicts the temperature-field in the tissue for different values of time derivative order $\alpha$ and of the spectral dimension $d_s$ 
(where $D_f=1.6$, the average value obtained from histological cuts). As before, we observe that for increasing values of $\alpha$, 
the thermal response of the tissue follows the same three different regimes as mentioned before: damped, critically damped and under-damped. Yet, contrary to $D_f$, 
the spectral dimension only affects the apparent period and the amplitude of the oscillations in the converse way: 
the lower the spectral dimension, the shorter the apparent period and the lower the amplitude of the oscillations. This trend is noticeable only for larger values of $\alpha$.   
Taken together, the two panels of figure~\ref{fig:6} provide information on how a thermal stimulus on a tissue will decay away and how this decay is modified by different values of
$\alpha$ and the two characteristics $D_f$ and $d_s$ of the fractality of the tissue. 

Having looked at the time-dependence of temperature-field, we now consider its spatial distribution. 
Figures~\ref{fig:4} and~\ref{fig:3} allow to visualize the time evolution of spatial temperature profiles inside the tissue for different times, where we focus on the under-damped regime $\alpha=1.9$. 
We refer to appendix~C for curves with $\alpha=0.7$. Figure~\ref{fig:4} provides spatial profiles for a fixed value of $d_s$. 
Then we observe the oscillating thermal response of the tissue strongly varies with $D_f$ such that it is possible that the temperature falls below the homeostatic temperature. 
As expected from figure \ref{fig:6}, the amplitudes of the oscillations decrease in time. 
These behaviours are in contrast with the sub-diffusive regimes for which the core region of the tissue is always heated and the temperature slowly decays towards the homeostatic body temperature. 
Similarly, when $D_f$ is fixed, see figure~\ref{fig:3}, temperature oscillations also show large sensitivities with respect to the spectral dimension, 
with cycles of hot-cold thermal stresses for the treated tissue.

\section{Concluding remarks} \label{sec4}


In this work, we explored the thermal response of a biological tissue with a fractal structure and a long-term memory effect. 
The analytical solutions have been determined for a parabolic initial temperature profile and homeostatic boundary conditions. 
Our analysis revealed the sensitivity of bio-heat with respect to the spectral dimension $d_s$, a parameter that -to our knowledge- has not been identified for tumors. 
We also derived an inequality (\ref{ineq}) that sets a lower bound on the fractal dimension. The simulation results showed three major trends. 
First, for increasing values of the fractional derivation order $\alpha$, one observes three qualitatively different temporal evolution regimes for the temperature profile: 
a first sub-diffusive one without oscillations and reminiscent of the over-damped regime in the damped oscillator, 
a second one with a single local minimum of the temperature-field, also without further oscillations, 
and a third regime where the temporal evolution of the temperature-field shows quasiperiodic oscillations. 
Second, the quasiperiodic nature of the oscillatory regime is found to depend on the fractal nature of the medium through at least two non-trivial and independent characteristics: 
the fractal dimension $D_f$ and the spectral dimension $d_s$. Third, the fractal geometry of the tissue, e.g. $D_f$ (and hence the grade of the cancer) 
impacts the period of the oscillations and their amplitudes (the period increases when $D_f$ decreases, the amplitude maxima behaves the same way), 
whereas the topology of the tissue, e.g. $d_s$, affects the period of the oscillations and their amplitudes in the converse way (both increase when $d_s$ increases).  

Yet, there is still a long way to go before efficient hyperthermia protocols can be extracted from this model. 
A first working prospect is connected to the model we chose to account for heat propagation in tissues. 
The dominating approach indeed consists in fractionalizing Pennes bio-heat equation (\ref{pennes}), but this latter has many well-known physical shortcomings 
(infinite speed of the heat flux, thermal equilibration in blood vessels, unidirectional blood flow, uniform metabolic heating and thermal conductivity). 
Hence, alternate fractional models based on Wulff-Klinger, Chen-Holmes, Weinbaum-Jiji or Maxwell-Cattaneo equations (see \cite{khanafer2009synthesis,hristov2019bio} 
for an extensive review of the different existing formulations) need to be investigated for comparison with the present approach.
A second avenue for future developments is related to the memory effect. Indeed, the half-lifetime of HSPs being around 48 hours, implies that thermoresistance is a slow-kinetic phenomenon with a lag phase, 
typically of a few hours, followed by a peak about a day after heating, and a total decay after a week \cite{horsman2007hyperthermia} 
(the exact characteristic times depend of course on the cell type, the heating protocol). 
Therefore the power-law memory kernel associated to Caputo derivative may lack accuracy to account for realistic thermo-tolerance kinetics.

Finally, a scientific challenge comes from the experimental data on fractal and spectral  dimensions. 
Indeed, as illustrated by bone imaging \cite{akkari2008relations}, there is in general no monotonous relation between the fractal dimension of a volume 
(e.g. that of the tumor to be heated) and the fractal dimension of its two-dimensional projection 
(e.g. that of the histological cut). Hopefully, this state of affairs may entice the advent of new in vivo methods for 3D extraction of the fractal dimension and the spectral dimension.\\


{\it {Acknowledgments}--} 
AL thanks  the LPCT team for the hospitality, where most of this work was developed. 
MH was supported by the french ANR-PRME UNIOPEN (ANR-22-CE30-0004-01).

\newpage


\appsection{A}{Mathematical derivations}
We provide the details of the analytical calculations of the temperature-profile $T(t,r)$ analysed in the main text. 
To begin, we show that the chosen initial conditions lead to a case distinction between $\beta=1$ and $\beta>2$. 
Then their respective solutions will be derived separately. 

\subsection{Case distinction}
First, we analyse the solutions to the radial equation, see eq.~(\ref{sep-space}) 
\begin{equation}  \label{Tspace-part} 
r^{2}z''(r)+(\beta+2)r z'(r)+\beta z(r)=-(\omega^{\alpha}-\gamma)r^{\beta+D_f-1}z 
\end{equation}
Introducing $\gamma_{\alpha} :=\omega^\alpha-\gamma$ and using from (\ref{gl:beta}) that $\beta+D_f-1=2+\theta$, we get in turn
\begin{equation}
r^{2}z''(r)+(\beta+2)r z'(r)+(\beta+\gamma_\alpha r^{(\theta+2)})z(r)=0, \label{space-part}
\end{equation}
\begin{subequations} 
An overview of the possible solutions of this leads to the following. 
In the first case, when $\gamma_\alpha=0$, eq.~(\ref{Tspace-part}) is homogeneous. Then a simple power-law ansatz, for $\beta\neq 1$, gives the two solutions
\BEQ
z(r) = C_1 r^{-1} + C_2 r^{-\beta}
\EEQ
where $C_{1,2}$ are constants to be fixed from the initial conditions. If $\beta=1$, one rather has
\BEQ
z(r) = C_1 r^{-1} + C_2 r^{-1} \ln r
\EEQ
\end{subequations} 

In the second case, when $\gamma_\alpha\neq 0$, eq.~(\ref{Tspace-part}) is inhomogeneous. 
Then the solution is found by adapting Frobenius' method \cite{Boas1983} 
and can be written in terms of Bessel functions $J,Y$ \cite{Abra65} as 
\begin{subequations} \label{eqn12}
\begin{align} 
z(r) &= C_1 r^{-\frac{\beta+1}{2}} J_{(\beta-1)/(\theta+2)}\left(\frac{2\sqrt{\gamma_{\alpha}}}{(\theta+2)} r^{(\theta+2)/2}\right) 
+  C_2 r^{-\frac{\beta+1}{2}} Y_{(\beta-1)/(\theta+2)}\left(\frac{2\sqrt{\gamma_{\alpha}}}{(\theta+2)} r^{(\theta+2)/2}\right) \label{eqn12a} \\
&= D_1 r^{-\frac{\beta+1}{2}} J_{(\beta-1)/(\theta+2)}\left(\frac{2\sqrt{\gamma_{\alpha}}}{(\theta+2)} r^{(\theta+2)/2}\right) 
+  D_2 r^{-\frac{\beta+1}{2}} J_{-(\beta-1)/(\theta+2)}\left(\frac{2\sqrt{\gamma_{\alpha}}}{(\theta+2)} r^{(\theta+2)/2}\right) \label{eqn12b} 
\end{align}
\end{subequations} 
\noindent where $C_{1,2}$ or $D_{1,2}$ are constants which are to be fixed from boundary conditions and in (\ref{eqn12b}) 
$\nu := \frac{\beta-1}{\theta+2}$ is required to be not an integer.  

Now, we use the series expansion \cite[(9.1.10)]{Abra65} of the Bessel function $J_{\nu}(x)$  to derive the leading $r\to 0$ behaviour in (\ref{eq13}), using (\ref{eqn12b}) 
\begin{eqnarray}
g(r) &\simeq&  D_1 r^{(\beta-1)/2} \left( \frac{\gamma_{\alpha}^{1/2}}{\theta+2} r^{(\theta+2)/2}\right)^{(\beta-1)/(\theta+2)} 
\left( 1 - \frac{1}{\Gamma\bigl(\frac{\beta-1}{\theta+2}+2\bigr)}  
\left( \frac{\gamma_{\alpha}^{1/2}}{\theta+2} r^{(\theta+2)/2}\right)^2+ {\rm O}\bigl(r^{2(\theta+2)} \bigr)\right) 
\nonumber \\
& & + D_2 r^{(\beta-1)/2} \left( \frac{\gamma_{\alpha}^{1/2}}{\theta+2} r^{(\theta+2)/2}\right)^{-(\beta-1)/(\theta+2)} 
\left( 1 - \frac{1}{\Gamma\bigl(-\frac{\beta-1}{\theta+2}+2\bigr)}  
\left( \frac{\gamma_{\alpha}^{1/2}}{\theta+2} r^{(\theta+2)/2}\right)^2+ {\rm O}\bigl(r^{2(\theta+2)}\bigr)\right)
\nonumber \\
&=& D_1 \left( \frac{\gamma_{\alpha}^{1/2}}{\theta+2}\right)^{(\beta-1)/{(\theta+2)}} r^{\frac{\beta-1}{2} 
+ \frac{\beta-1}{2}} \bigl(1 + {\rm O}(r^{\theta+2}) \bigr) 
\nonumber \\
& & + D_2 \left( \frac{\gamma_{\alpha}^{1/2}}{\theta+2}\right)^{-(\beta-1)/{(\theta+2)}} r^{\frac{\beta-1}{2} - \frac{\beta-1}{2}} 
\left( 1 - \frac{\bigl( \gamma_{\alpha}^{1/2}/(\theta+2)\bigr)^2}{\Gamma\bigl( 2 - \frac{\beta-1}{\theta+2}\bigr)} r^{\theta+2} 
+ {\rm O}(r^{2(\theta+2)}) \right)
\nonumber \\
&=& E_2 \cdot 1 + E_1 r^{\beta-1} - E_2' r^{\theta+2} + \ldots 
\label{gl:asymp}
\end{eqnarray}
and where $E_1,E_2,E_2'$ are known constants. Since we always have from (\ref{gl:beta}) that $\beta-1 < \theta+2$, and recall from (\ref{gl:racc}) that $\theta>0$, 
the second term will for $r\to 0$ always dominate over against the third one.

But we also require to have differentiable solutions when $r\to 0$. Then the parameter region $0<\beta-1<1$ 
must be excluded, since this would a non-differentiable solution at $r=0$ which would contradict (\ref{init2}). 
Then it is only possible that either $\beta=1$ or else $\beta-1>1$ which means $\beta>2$. 
This conclusion is not changed by the third term. 

\subsection{Case $\beta=1$}

Although we shall argue in the main text that this case looks unphysical since it leads to $D_f>2$, the derivation of the solution is instructive enough to be outlined here. 
Since in this case, the two first terms in (\ref{gl:asymp}) are confounded, it is here better to use the form (\ref{eqn12a}) 
and the unwanted logarithmic singularity of $Y_0$ at $r=0$ implies that we must take $C_2=0$. Using also the initial condition (\ref{init2-bis}), 
we find that the temperature-field becomes, for $0<\alpha\leq 2$   
\begin{eqnarray}
    T(t,r)=T_0+ \int \!\D\omega\:  c_{\omega}\text{E}_{\alpha}\left(- \omega^\alpha t^{\alpha}\right)\:  
    J_{0}\left(\frac{2\sqrt{\gamma_{\alpha}}}{D_f} r^{D_f/2}\right)  \label{part-sol}
\end{eqnarray}
The first boundary condition (\ref{init2}) is fulfilled when the argument of the 0$^{\rm th}$-order Bessel function obeys 
\begin{eqnarray}
  \omega_n=\left(\mathcal{D}\gamma+\mathcal{D}\frac{\kappa^2_{0,n}D_f^2}{4R^{D_f}}\right)^{1/\alpha}
\end{eqnarray}
Here, $\left\{\kappa_{0,n},n\in\mathbb{N}\right\}$ is the set of discrete zeros of $J_0$ \cite{Abra65}. Hence, the temperature field is 
\begin{equation} \label{gl:T-beta1}
    T(t,r)= T_0+ \sum_{n=1}^{\infty} c_n\text{E}_{\alpha}\left(-\ \left[\mathcal{D}\gamma+\mathcal{D}\frac{\kappa^2_{0,n}D_f^2}{4R^{D_f}}\right] t^{\alpha}\right)\:  
    J_{0}\left(\kappa_{0,n} \left(\frac{r}{R}\right)^{D_f/2}\right) 
\end{equation}
To evaluate the $c_n$, one uses the initial condition (\ref{init2}) and the orthogonality property of the Bessel functions. 

The normalisation can be easily expressed in terms of the dimensionless variable $x=\sqrt{(r/R)^{D_f}}$, 
such that we can use the orthogonality relation \cite[eq. (19.10)]{Boas1983} 
\begin{equation}
\int_0^1 \D x\: xJ_0(\kappa_{0,n}x)J_0(\kappa_{0,m}x) = \demi {J_1^2(\kappa_{0,n})}\: \delta_{nm}.
\end{equation}
The coefficients of the linear combination are given by, see \cite[(2.12.1.1)]{Prudnikov1,Prudnikov2}
\BEA
c_n &=&\frac{2(T_c-T_0)}{J_1(\kappa_{0,n})^2}\int_0^1 \D x\: x[(1-x^{4/D_f})J_0(\kappa_{0,n}x)] \nonumber \\
    &=& \frac{2(T_c-T_0)}{J_1(\kappa_{0,n})^2}\left[\frac{J_1(\kappa_{0,n})}{\kappa_{0,n}}
-\frac{D_f}{2(D_f+2)}{_1}F_{2}\left(1+\frac{2}{D_f};1,2+\frac{2}{D_f};-\frac{1}{4}{\kappa_{0,n}^2}\right)\right],  
\label{gl:cn-beta1}
\EEA
where $_{1}F_{2}$ is a generalised hyper-geometric function. Eqs.~(\ref{gl:T-beta1},\ref{gl:cn-beta1}) together give the sought solution.


\subsection{Case $\beta>2$}
This case is histologically of interest since $D_f<2$. 

The first boundary condition (\ref{init2}) is now fulfilled by choosing $D_1$ and $D_2$ such that
\begin{eqnarray}\label{eq23}
    \lefteqn{T(t,r)=T_0+\int \!\D\omega\: c_{\omega}\:  \text{E}_{\alpha}\left(-\omega^\alpha t^{\alpha}\right)r^{\frac{\beta-1}{2}} \times} \nonumber \\
    &\times& \left[J_{-(\beta-1)/(\theta+2)}\left(\frac{2\sqrt{\gamma_{\alpha}}}{(\theta+2)} R^{(\theta+2)/2}\right)  
    J_{(\beta-1)/(\theta+2)}\left(\frac{2\sqrt{\gamma_{\alpha}}}{(\theta+2)} r^{(\theta+2)/2}\right) \right. \nonumber \\
    & & \left. - J_{(\beta-1)/(\theta+2)}\left(\frac{2\sqrt{\gamma_{\alpha}}}{(\theta+2)} R^{(\theta+2)/2}\right)
               J_{-(\beta-1)/(\theta+2)}\left(\frac{2\sqrt{\gamma_{\alpha}}}{(\theta+2)} r^{(\theta+2)/2}\right)\right] \label{sol-b2-partial}
\end{eqnarray}  
which guarantees that the square bracket vanishes for $r=R$. 
We now turn to the third condition (\ref{init2}). The $r$-dependent term in the first line of the square bracket in (\ref{eq23}) is proportional to
$x^{\nu}J_{\nu}(x)$, where $0<\nu=\frac{\beta-1}{\theta+2}<1$, and increases with $x$ for $x$ small enough. On the other hand, the $r$-dependent term in the second line
of the square bracket is proportional to $x^{\nu}J_{-\nu}(x)$ which for $x$ small enough decreases from a constant value at $x=0$.  
The leading terms in (\ref{gl:asymp}) tells us that the first line of the square bracket in (\ref{eq23}) will always dominate over against the second line in the 
$r\to 0$ limit, whatever the choice of the coefficients. 
Hence any linear combination of the two terms in (\ref{eq23}) will be dominated by the term in the first line, if that one is present at  
all, and for $x\ll 1$ should always increase with $x$. Therefore, it {\em cannot} have a local maximum at $r=x=0$. 
The only exit from this dilemma is that the $R$-dependent factor in the first line vanishes, which means that 
\begin{equation}
    \frac{2\sqrt{\gamma_{\alpha}}}{(\theta+2)} R^{(\theta+2)/2} \stackrel{!}{=} \kappa_{-\nu,n} ~~\Longrightarrow~~
\omega^\alpha_n=\mathcal{D}\gamma+\mathcal{D}\left(\frac{\kappa_{-\nu,n}(\theta+2)}{2R^{(\theta+2)}}\right)^2
\label{sign-arg}
\end{equation}
where $\kappa_{-\nu,n}$ is the $n^{\rm th}$ zero of the Bessel function $J_{-\nu}(x)$ such that $J_{-\nu}(\kappa_{-\nu,n})=0$. \cite{Abra65}. 

Furthermore, the definition (\ref{gl:beta}) translates for $\beta>2$ to $\theta>D_f-1$. 
Since $\theta$ is not independent from the fractal dimension, using its definition $\theta+2=2D_f/d_s$ in eq.~(\ref{gl:racc}), 
we finally obtain a condition among the fractal and spectral dimensions, namely 
$D_f>d_s/(2-d_s)$. We can then explore the relevant physically acceptable solutions in terms of these dimensional quantities.
To find the expansion coefficients, we set $x=(r/R)^{(\theta+2)/2}$ 
such that the temperature profile can be recast as
\begin{eqnarray}
    T(t,x) = T_0+R^{(\beta-1)/2}x^\nu\sum_{n=1}^\infty \: \, c_{\nu,n}\,\text{E}_\alpha(-\omega_n^\alpha t^\alpha) 
     J_{-\nu}(\kappa_{-\nu,n} x).
\end{eqnarray}
Then, the assumed initial parabolic temperature profile 
\begin{equation}
T(0,r)=T_0+(T_c-T_0)\left[1-\left(\frac{r}{R}\right)^2\right]    
\end{equation}
becomes
\begin{equation}
  T(0,x) = T_0+(T_c-T_0)(1-x^{\frac{4}{\theta+2}}).  
\end{equation}
Therefore, upon using the orthogonality relation for the Bessel functions \cite[eq. (19.10)]{Boas1983}
\begin{equation}
\int_0^1 \D x\: xJ_{\mu}(\kappa_{\mu,n}x)J_\mu(\kappa_{\mu,m}x) = \demi {J_{\mu+1}^2(\kappa_{\mu,n})}\: \delta_{nm},
\end{equation}
the expansion coefficients are obtained through
\begin{equation}
c_{\nu,n}=\frac{2(T_c-T_0)}{R^{(\beta-1)/2}J^2_{-\nu+1}(\kappa_{-\nu,n})}\int_0^1 \!\D x\: x^{1-\nu}\left(1-x^{4/(\theta+2)}\right)J_{-\nu}(\kappa_{-\nu,n}x).    
\end{equation}
Using {\it Mathematica}, or eqs. (1.8.1.1) and (1.8.1.21) in \cite{Prudnikov2}, 
one gets the result
\begin{equation}
\int_0^1 \!\D x\: \left(1-x^s\right) x^{1-\nu } J_{-\nu }(b x) 
= \frac{J_{1-\nu }(b)}{b}-\frac{2^{\nu } b^{-\nu } \, _1F_2\left(\frac{s}{2}-\nu +1;1-\nu ,\frac{s}{2}-\nu +2;-\frac{b^2}{4}\right)}{(-2 \nu +s+2)\Gamma(1-\nu )},   
\end{equation}
with the generalised hypergeometric function $_1F_2(a;b;z)$.

This solution is valid if $b\in\mathbb{R}_+$ and $\nu<1$. 
Then, the expansion coefficients read  
\begin{equation} \label{gl:31}
c_{\nu,n}=\frac{2(T_c-T_0)}{R^{(\beta-1)/2}J^2_{1-\nu}(\kappa_{-\nu,n})}\left[\frac{J_{1-\nu }(\kappa_{-\nu,n})}{\kappa_{-\nu,n}}
-\frac{2^{\nu } \kappa_{-\nu,n}^{-\nu } \, 
_1F_2\left(\frac{s}{2}-\nu +1;1-\nu ,\frac{s}{2}-\nu +2;-\frac{\kappa_{-\nu,n}^2}{4}\right)}{(-2 \nu +s+2)\Gamma(1-\nu )}\right] 
=: \frac{d_{\nu,n}}{R^{(\beta-1)/2}},   
\end{equation}
with $s=4/(\theta+2)$. We stress that the set of parameters $\{\beta,\nu,s\}$ are not independent since by its definition $\nu=1-D_f/(\theta+2)$, 
whereas the restriction $\beta\ge2$ implies in turn that $\theta+2\ge D_f-1.$ 
Thus, once the value of the fractal dimension is set, this fixes the allowed frequencies, which in turn determines the suitable fractional order of the time derivative 
$\alpha$. 
Therefore, the temperature profile in this case, satisfying all initial and boundary conditions (\ref{init2-total}), reads for $0<\alpha\leq 2$
\begin{eqnarray} \label{gl:32}
    T(t,r) = T_0+\left(\frac{r}{R}\right)^{(\theta+2-D_f)/2}\sum_{n=1}^\infty \: \, d_{1-D_f/(\theta+2),n}\,\text{E}_\alpha(-\omega_n^\alpha t^\alpha) 
     J_{D_f/(\theta+2)-1}\left(\kappa_{D_f/(\theta+2)-1,n}\left(\frac{r}{R}\right)^{(\theta+2)/2}\right)\nonumber.\\
\end{eqnarray}
Eqs.~(\ref{gl:31},\ref{gl:32}) give the sought solution and are (\ref{glT:31},\ref{solutionT}) in section~\ref{sec:3} in the main text. 

\appsection{B}{Fractional calculus}
We provide some background on the fractional Caputo derivative, fractional differential equations and the Mittag-Leffler functions. 
This will be used throughout in the main text. We base our presentation on \cite{Diethelm2010,Gorenflo,podlubny1998fractional}. 

For $0<\eta< 1$, the {\em Caputo derivative} of order $\alpha=m+\eta$ with $m\in\mathbb{N}$ is defined for $m+1$ times continuously differentiable functions 
$f\in C^{m+1}(0,\infty)$ as
\BEQ \label{Dc}
{^c}D^{\alpha} f(x) = {^c}D^{m+\eta} f(x) := \frac{1}{\Gamma(1-\eta)}\int_0^{x} \!\D t\: \frac{f^{(m+1)}(t)}{(x-t)^{\eta}} 
= \frac{1}{\Gamma( \lceil\alpha\rceil - \alpha)} \int_0^{x} \!\D t\: \frac{f^{(\lceil\alpha\rceil)}(t)}{(x-t)^{\alpha+1 -\lceil\alpha\rceil}} 
\EEQ
where $f^{(m+1)}(t) = \frac{\D^{m+1} f(t)}{\D t^{m+1}}$ is the ordinary derivative of integer order $m+1\geq 1$. In addition $\lceil\alpha\rceil$ is the smallest integer
larger than or equal to $\alpha$ (also known as Gau{\ss}'s bracket). Obviously, one has $^cD^0 f(x)=f(x)$ and $^cD^n f(x) = f^{(n)}(x)$ when $n\in\mathbb{N}$. 
As an example, the Caputo derivative of order $0<\eta<1$ of a simple power $f(x)=x^\beta$ is 
\begin{equation}\label{power}
^{c}D^\eta x^\beta=\frac{\Gamma(\beta+1)}{\Gamma(\beta+1-\eta)}x^{\beta-\eta} \;\;,\;\;\beta\neq0
 \end{equation}
whereas for a constant function (i.e. $\beta=0$), the Caputo definition (\ref{Dc}) leads to the result 
\begin{equation}\label{constant}
^{c}D^\eta \,1=0.
 \end{equation}
familiar for a standard derivative. For a linear combination of functions $f(x)=a_1f_1(x)+f_2(x)$, the Caputo derivative is clearly a linear operator 
\begin{equation}\label{power2}
{^c}D^\eta f(x)=a_1 {^c}D^\eta f_1(x)+a_2 {^c}D^\eta f_2(x).
 \end{equation}
The property (\ref{constant}) is suggestive and motivates our choice of the Caputo fractional derivative in this work, although there exist by now alternative
definitions of inequivalent fractional derivatives which share the property (\ref{constant}) \cite{Caputo2015}. 
Using the Caputo derivative appears to us as the most suitable one to describe the physical scenario of interest, 
because of (\ref{constant}) initial values can be specified in terms of values of the function
$f(x)$ at a specified points (and if necessary of its ordinary derivatives). On the other hand, the composition of Caputo derivatives is more complicated
than for the standard one. Only if there is some $\ell\in\mathbb{N}$ such that {\em both} $\alpha, \alpha+\vep\in[\ell-1,\ell]$ 
are in the same sub-interval, then one has
\BEQ
{^c}D^{\vep} \bigl( {^c}D^{\alpha} f(x)\bigr)  = {^c}D^{\alpha+\vep} f(x)
\EEQ
but there are counter-examples which show this identity to be wrong when the condition $\alpha, \alpha+\vep\in[\ell-1,\ell]$ is not met \cite[p. 57]{Diethelm2010}. 
Also, the Caputo derivative is in general {\em not} commutative. 

\begin{figure}[tb]
\begin{center}
\includegraphics[width=.46\hsize]{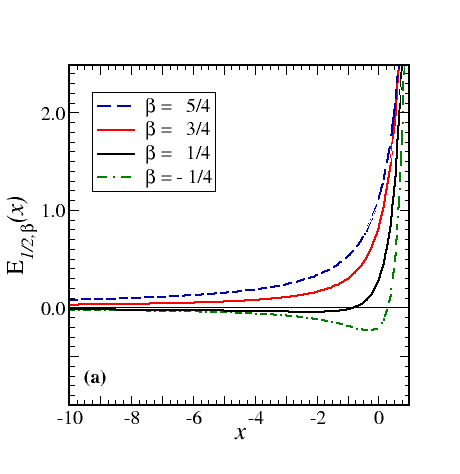} ~~~ \includegraphics[width=.46\hsize]{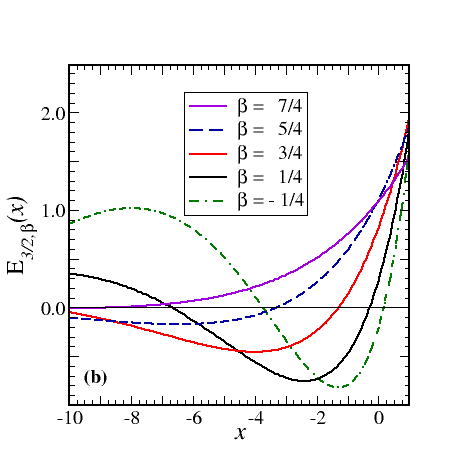}
\end{center}
\caption[figB1]{\small Plots of the generalised Mittag-Leffler function $\text{E}_{\alpha,\beta}(x)$ for {\bf (a)} $\alpha=\demi$ and {\bf (b)} $\alpha=\frac{3}{2}$ 
and several values of $\beta$ as indicated. The line $\text{E}_{\alpha,\beta}(x)=0$ is indicated by the thin full line. 
\label{figB1} }
\end{figure}

An important application concerns differential equations. For example, the following initial-value problem
\BEQ \label{gl:dgl1} 
{^c}D^{\alpha} y(x) = \lambda y(x) \;\; , \;\; y(0) = y_0 \;\; , \;\; y^{(k)}(0) = 0 \mbox{\rm\small ~~~for $k=1,2,\ldots,\lceil\alpha\rceil-1$}
\EEQ
with a constant $\lambda$ has the unique solution
\BEQ \label{gl:sol1}
y(x) = y_0 \text{E}_{\alpha}\bigl(\lambda x^\alpha\bigr)
\EEQ 
where $\text{E}_{\alpha}(x) = \sum_{k=0}^{\infty} \frac{x^k}{\Gamma(\alpha k+1)}$ is the Mittag-Leffler function \cite{Gorenflo}. From the context of the initial-value problem 
(\ref{gl:dgl1}), one sees
that $\text{E}_{\alpha}(x)$ is a natural generalisation of the exponential function to the situation of fractional derivatives, see also the examples (\ref{gl:exem}) below. 
We shall also need the generalised Mittag-Leffler function \cite{Gorenflo}
\BEQ \label{gl:MittagLeffler2}
\text{E}_{\alpha,\beta}(x) = \sum_{k=0}^{\infty} \frac{x^k}{\Gamma(\alpha k+\beta)}
\EEQ
and trivially $\text{E}_{\alpha,1}(x)=\text{E}_{\alpha}(x)$. Elementary special cases include
\BEQ \label{gl:exem} 
\text{E}_{1,1}(x) = \text{E}_{1}(x) = e^x \;\; , \;\; 
\text{E}_{1,2}(x) = \frac{e^x-1}{x} \;\; , \;\;
\text{E}_{2,1}(x) = \text{E}_{2}(x) = \cosh(\sqrt{x\,}\,) \;\; , \;\;
\text{E}_{2,2}(x) = \frac{\sinh(\sqrt{x\,}\,)}{\sqrt{x\,}}
\EEQ
Sometimes it is useful to relate this to generalised hypergeometric functions \cite{Prudnikov3}
\BEA
\text{E}_{1,\beta}(x)  &=& \frac{1}{\Gamma(\beta)} {}_1F_{1}\bigl(1;\beta;x\bigr) \nonumber \\
\text{E}_{2,2\beta}(x) &=& \frac{1}{\Gamma(2\beta)} {}_1F_{2}\bigl(1;\beta,\beta+\demi;\frac{x}{4}\bigr)  \\
\text{E}_{3,3\beta}(x) &=& \frac{1}{\Gamma(3\beta)} {}_1F_{3}\bigl(1;\beta,\beta+\frac{1}{3},\beta+\frac{2}{3};\frac{x}{27}\bigr) \nonumber 
\EEA
which allows to express more simply
\BEQ
\text{E}_{\demi,\beta}(\pm x) = \text{E}_{1,\beta}\bigl(x^2\bigr) \pm x \text{E}_{1,\beta+\demi}\bigl( x^2\bigr) \;\; , \;\;
\text{E}_{\frac{3}{2},\beta}(\pm x) = \text{E}_{3,\beta}\bigl(x^2\bigr) \pm x \text{E}_{3,\beta+\frac{3}{2}}\bigl( x^2\bigr)
\EEQ
Some examples are displayed in figure~\ref{figB1} (curves of $E_{\alpha}(-x^{\alpha})$ can for example be found in 
\cite{Diethelm2010,manapany2024,Gorenflo}). For large positive $x$, one sees the rapid exponential increase with $x$; for large negative $x$, the decay
is considerable more slow and is modulated by oscillations. 

Integral representations for the efficient numerical computation of $\text{E}_{\alpha,\beta}(x)$ are given in 
\cite{Hilfer2006,Parovik2012,Saenko2020}.\footnote{Warning: the 
english translation \cite{Parovik2016} contains quite a few typos not present in the russian original \cite{Parovik2012}.} 
For example, if $0<\alpha\leq 2$, the fractional differential equation
\BEQ \label{gl:dgl} 
{^c}D^{\alpha} y(x) = \lambda y(x)
\EEQ
has the general solution
\BEQ  \label{gl:sol2}
y(x) = C_1 \, \text{E}_{\alpha,1}(\lambda x^{\alpha}) + C_2 \, x\, \text{E}_{\alpha,2}(\lambda x^{\alpha})
\EEQ
where the constants $C_{1,2}$ will be fixed by the initial conditions associated with the differential equation (\ref{gl:dgl}). 
A lot of mathematical analysis on the smoothness and other properties of the solutions
of such differential equations is available \cite{dias2005effects}. On the other hand the exponential $e^{\lambda x}$ 
is {\em not} an eigenfunction of the Caputo operator, 
since that one is given by (\ref{gl:sol1}). Rather, one has by generalising (\ref{power}) and using the linearity of ${^c}D^{\alpha}$
\BEQ
{^c}D^{\alpha} e^{\lambda x} = \sum_{k=0}^{\infty} \frac{\lambda^{k+\lceil \alpha\rceil} t^{k+\lceil\alpha\rceil-\alpha}}{\Gamma(k+1+\lceil\alpha\rceil-\alpha)} 
= \lambda^{\lceil\alpha\rceil} t^{\lceil\alpha\rceil - \alpha} \text{E}_{1,\lceil\alpha\rceil-\alpha+1}(\lambda t)
\EEQ
where $\lambda$ is an arbitrary complex constant. Consequently, the Caputo fractional derivatives of the hyperbolic and trigonometric functions follow immediately. 

Linear differential equations with constant coefficients are often treated via a  La\-place-trans\-for\-ma\-tion
$\mathscr{L}(f)(p) := \int_0^{\infty} \!\D x\: e^{-px} f(x) = \lap{f}(p)$. One has
\BEQ
\mathscr{L}\bigl({^c}D^{\alpha} f\bigr)(p) = \bigl(\lap{ {^c}D^{\alpha} f}\,\bigr)(p) = p^{\alpha} \lap{f}(p) - \sum_{k=0}^{\lceil\alpha\rceil} p^{\alpha-k-1} f^{(k)}(0) 
\EEQ
Along with the identity $\mathscr{L}\bigl( x^{\beta-1} \text{E}_{\alpha,\beta}(\lambda x^{\alpha}) \big)(p) = \frac{p^{\alpha-\beta}}{p^{\alpha}-\lambda}$ 
which follows straightforwardly from the definition (\ref{gl:MittagLeffler2}) (with $p,\alpha,\beta$ positive), 
this provides  elementary and easy proofs of the statements (\ref{gl:sol1},\ref{gl:sol2}). 

These statements provide the basis for the solution of linear fractional partial differential equations via the separation-of-variables method used in the text, 
which generalises existing treatments of the simple diffusion equation \cite{manapany2024}. 

In the literature, another popular choice for a fractional derivative is the {\em Riemann-Liouville derivative}. For the order $0<\eta<1$, it is defined as
\begin{equation} \label{Dr}
^{r}D^\eta f(x):=\frac{1}{\Gamma(1-\eta)}\frac{\D}{\D x}\int_0^x \!\D y\: \frac{f(y)}{(x-y)^\eta}
\end{equation}
Obviously, that is another linear operator. Then it is related to the Caputo derivative by 
\begin{equation} \label{B17}
^{r}D^\eta f(x)={^c}D^\eta f(x)+\frac{f(0)}{\Gamma(1-\eta)x^\eta}.
\end{equation}
which implies that the Riemann-Liouville derivative of a constant ${^r}D^{\eta}\, 1 = \frac{x^{-\eta}}{\Gamma(1-\eta)}\neq 0$ does not vanish. 
This makes it less convenient for the kind of study we present. Extensions of (\ref{B17}) to higher orders are known \cite[p.53]{Diethelm2010}.  
Historically, the Riemann-Liouville arose, in the early 19$^{\rm th}$ century, as the first example of a `fractional derivative' since it allows to write
the solution of Abel's integral equation
\begin{equation}
g(x) =\int_0^{x}\!\D y\: \frac{f(y)}{(x-y)^\eta} 
\end{equation}
where $g(x)$ is assumed known and $f(x)$ is to be found, as a simple formal expression, namely 
\BEQ
f(x)=\frac{\sin\bigl(\pi \eta\bigr)\Gamma(1-\eta)}{\pi}\, \bigl({^r}D^{\eta} g\bigr)(x) 
= \frac{\sin\pi\eta}{\pi}\frac{\D}{\D x}\int_0^x \!\D y\: \frac{g(y)}{(x-y)^{1-\eta}}
\end{equation}
again for $0<\eta<1$. See \cite{Diethelm2010} for a full mathematical treatment. 

\appsection{C}{Further information}

\begin{figure}[h!]
    \centering
    \includegraphics[width=1\linewidth]{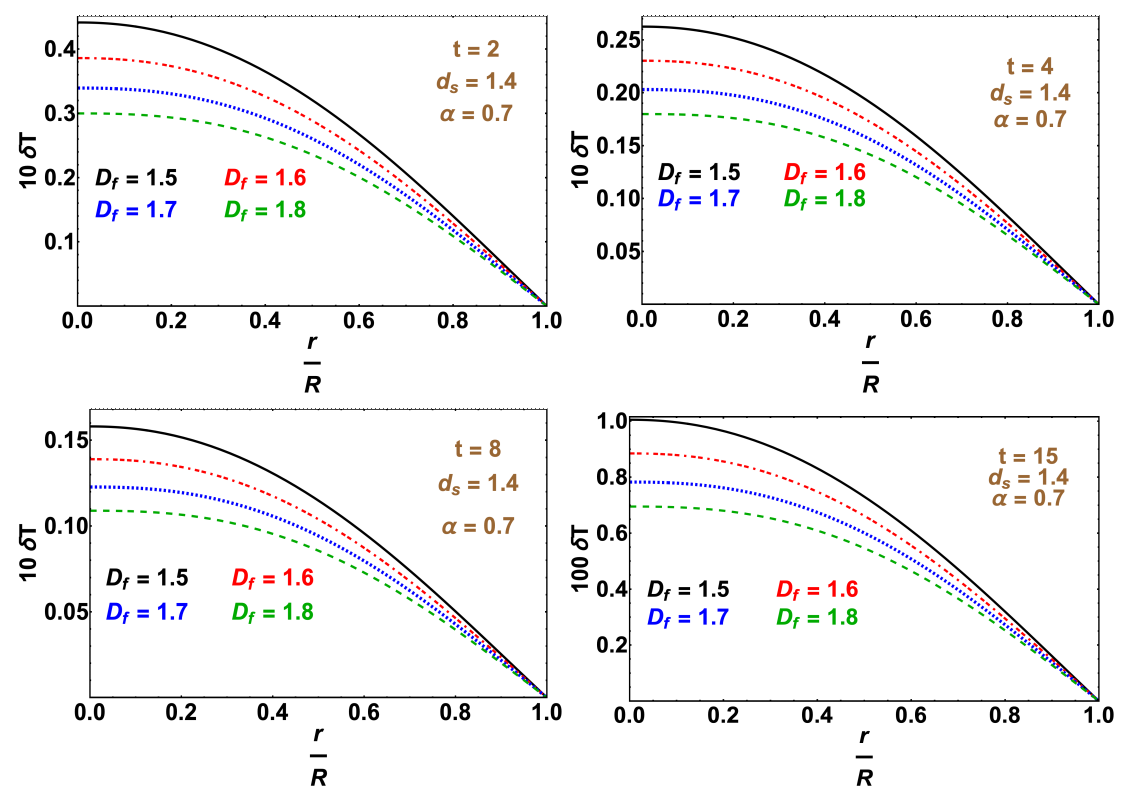}
    \caption{Snapshots of the normalized differential temperature profile versus normalized distance $r/R$, 
    at four selected times for $\alpha=0.7$ and $d_s=1.4$. The curves are parametrized by the fractal dimension $D_f$.} 
    \label{fig:C1}
\end{figure}
\begin{figure}[h!]
    \centering
    \includegraphics[width=0.495\linewidth]{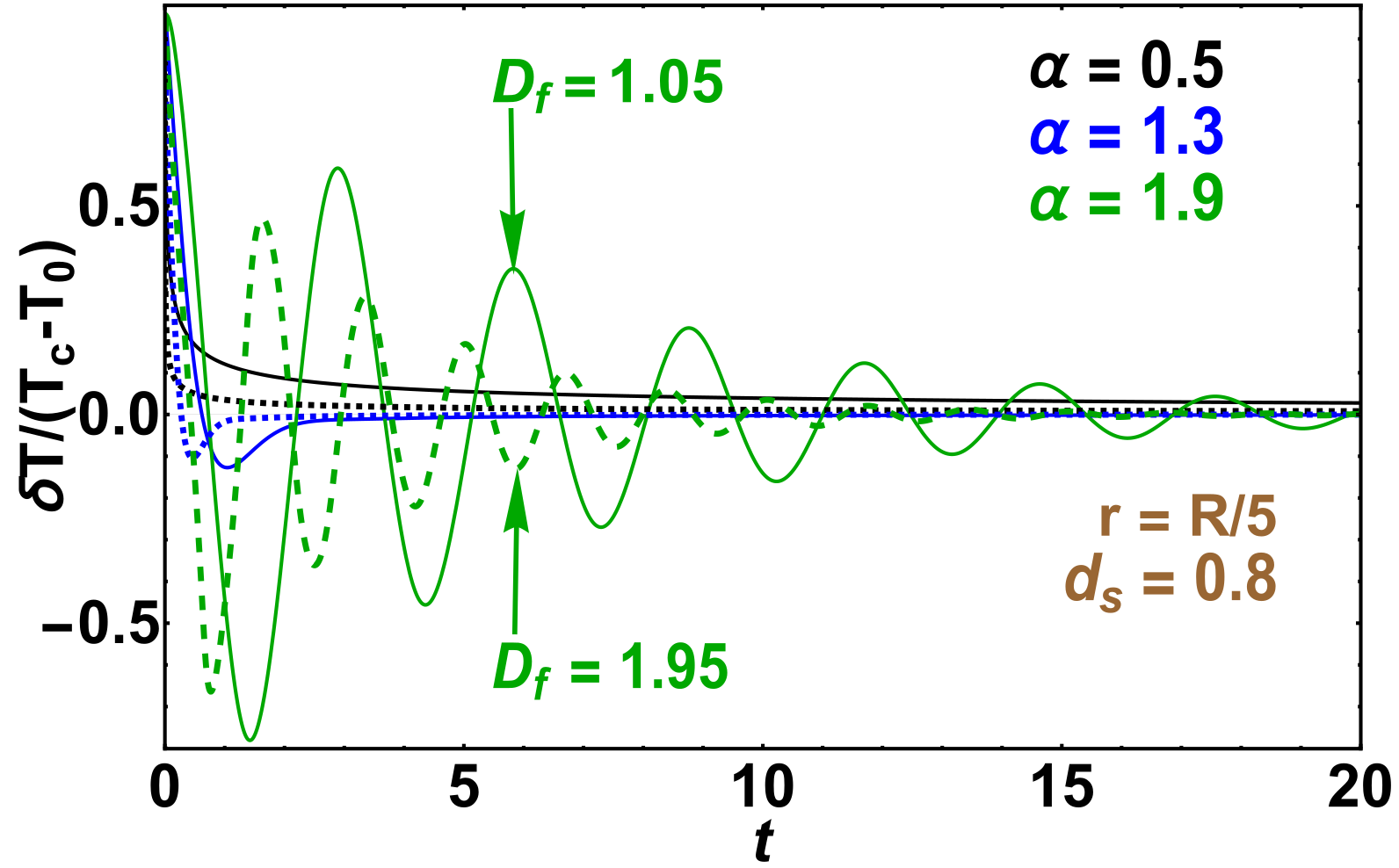}~\includegraphics[width=.495\linewidth]{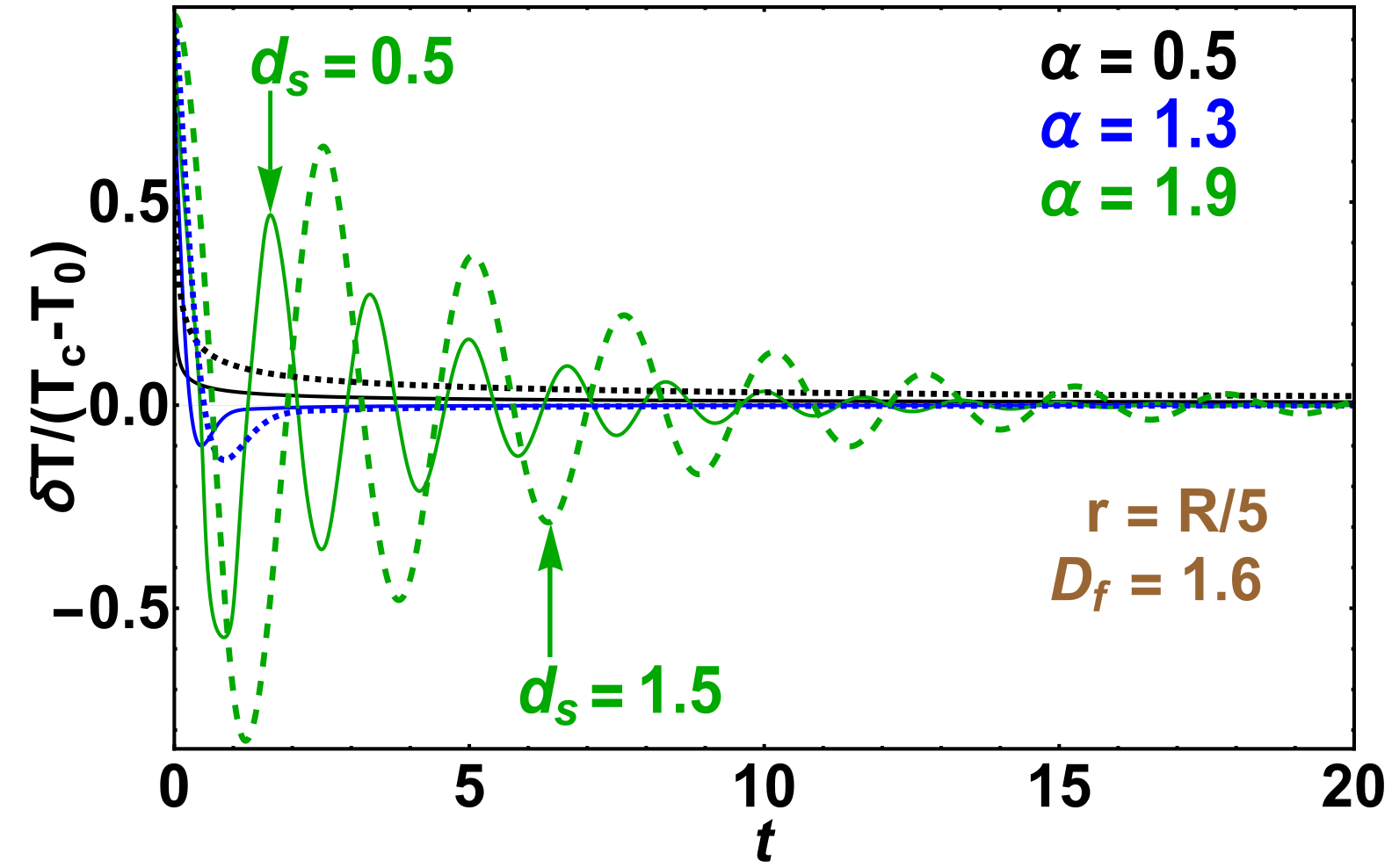}
    \caption{Time evolution of the normalised differential temperature at $r=0.2R$, at three selected values for $\alpha=0.5, 1.3$ and $1.9$. 
    The left panel corresponds to fixed spectral dimension and the thin (thick) line is parametrised by the fractal dimension $D_f=1.05$ or $D_f=1.95$ as shown by the arrows. 
    In the right panel, the fractal dimension is set to $D_f=1.6$ and the different lines correspond to spectral dimensions $d_s=0.5$ and $d_s=1.5$ as shown by the arrows.}
    \label{fig:6a}
\end{figure}
As a complement to the main text, we show in figure~\ref{fig:C1} the spatial profiles
in the sub-diffusive case $\alpha=0.7$ and a fixed value $d_s=1.4$. For all values of the 
fractal dimension $D_f$ considered, the profiles look very similarly and their amplitude
decays rapidly to zero. The initial order of the curves in maintained for all times. This 
is different with respect to what we found when $\alpha=1.9$, see figures~\ref{fig:4} and~\ref{fig:3}. 

In figure~\ref{fig:6a}, further data on the time-envelope of the temperature-field are shown, for values of $\alpha$, $D_f$ and $d_s$ not covered in the main text. 
Here we have used parameter values that highlight the main trends depicted in fig. \ref{fig:6}.




\end{document}